# One-stage hollow-core fiber-based compression of Yb:KGW lasers to the single-cycle regime

Z. Pi, A. B. Enders, D. Vaso, and E. Goulielmakis*

Institute of Physics, University of Rostock, Germany
*Corresponding author: e.goulielmakis@uni-rostock.de

**The generation of intense, waveform-controlled, single-cycle pulses based on Yb:KGW amplifiers is central to integrating these lasers with attosecond metrology and spectroscopy. Here, we demonstrate single-stage, multi-octave (~ 2.4 octaves) spectral broadening of Yb:KGW amplified pulses in a neon-pressurized hollow-core fiber (HCF) capillary and their compression to the single-cycle regime (1.1 cycles at 880 nm) using chirped mirrors. Utilizing Homochromatic Attosecond Streaking (HAS), we characterize the field waveforms of the generated pulses and demonstrate precise control of their carrier-envelope phase. Our results provide a simplified route to single-cycle pulse generation using Yb:KGW technology, previously possible only with Ti:Sapphire-based front ends. This work paves the way for advanced applications in attosecond science, strong-field physics, and spectroscopy.**

The advent of intense few-cycle [1–5], single-cycle [6, 7], and sub-cycle [8–10] optical laser pulses has revolutionized ultrafast science, enabling new insights into strong-field light-matter interactions, attosecond physics, and electron dynamics on their atomic and molecular timescales [7–9]. Ti:Sapphire lasers have long been the cornerstone of such applications. However, they are accompanied by several limitations, including modest repetition rates (< 10 kHz) and challenges in scaling their average power. In recent years, the development of Yb-doped solid-state lasers (e.g., Yb:KGW) enabled mJ-level pulse energies at high repetition rates, making them attractive for applications requiring high photon flux or demanding data acquisition statistics [11–22]. Yet, their utility in ultrafast experiments hinges on the compression of their intrinsically long pulses (~ 170 to 300 fs) to the few- or single-cycle regime.

Over the past decade, critical efforts have been made to overcome this bottleneck. These include spectral broadening in hollow-core fibers (HCFs) [23] filled with gases exhibiting high nonlinearity (e.g., argon and xenon), pressurizing the gases within [13, 14], and extending the capillary length to several meters [15, 16]. Moreover, multi-plate [18] or multi-pass [19] spectral broadening stages, followed by compression with chirped mirrors have been employed. A critical advancement in compressing of mJ-scale Yb:KGW pulses towards sub-2-cycle durations, relevant for attosecond science, was attained using cascaded, two-stage modules [20, 21]. More recently this concept has been extended by employing pressurized neon (Ne) in both capillaries of a compression module, enabling the generation of multi-octave pulses and their compression to the single-cycle regime using a light-field synthesizer [21].

Nevertheless, the quest for a more compact and versatile compression setup remains essential for widespread adoption of single-cycle pulses in more laboratories. Recently, mJ-level, carrier-envelope phase (CEP) stabilized Yb:KGW lasers with a pulse duration of < 90 fs and high repetition rates > 10 kHz have become available. However, even for the well-explored Ti:Sapphire technology, compression by a factor of ~ 30, required to attain single cycles, is challenging unless several stages are employed [24]. Yet, the recent development of pulse-broadening strategies using highly pressurized, high ionization potential gases [19], offers new motivation for revisiting these approaches.

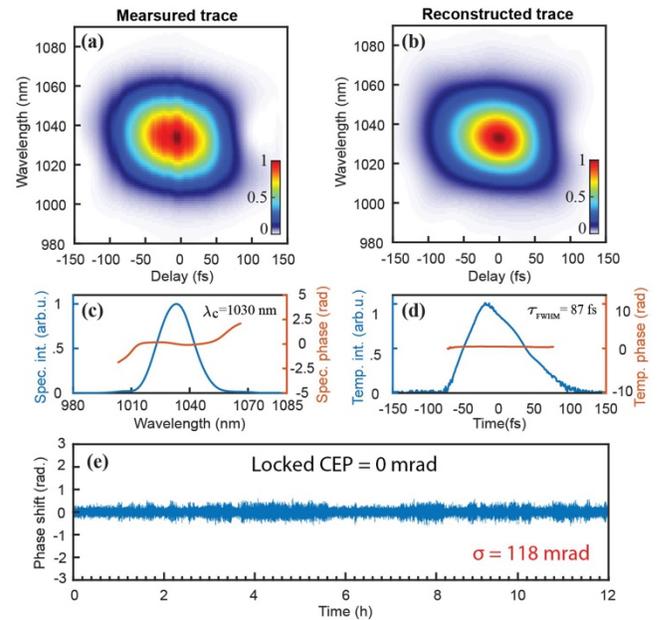

Fig. 1 (a) Measured TG-FROG trace. (b) Reconstructed TG-FROG trace. (c) Retrieved spectrum, and (d) temporal profile of the frond-end pulse. (e) Carrier-envelope phase (CEP) drift, locked at 0 rad over a period of 12 hours. The CEP stability has a standard deviation of 118 mrad.

The laser system used in this study is a commercial Yb:KGW regenerative amplifier (Pharos, Light Conversion), which delivers pulses with a duration of approximately 90 fs and an energy of ~ 1 mJ per pulse at a repetition rate of 10 kHz. The laser beam exhibits a high-quality Gaussian spatial profile with a $1/e^2$ diameter of 5.2 mm. Figs. 1(a)–(d) show pulse duration measurements of the laser front-end using a home-built transient grating frequency-resolved optical gating (TG-FROG) setup [25]. Our measurements yielded a duration of ~ 87 fs (FWHM), which is somewhat longer than the Fourier-transform-limited (FTL) pulse duration of 68 fs that can be inferred by its bandwidth of 24.8 nm (FWHM). Fig. 1(e) shows the CEP stability of the system recorded for a period of ~ 12 hours. Our experimental setup for spectral broadening and compression is schematically shown in Fig. 2 (a). To ensure that pulses of the shortest duration reach the entrance of the capillary, we adjusted the dispersion settings of the laser compressor to compensate for positive dispersion resulting from air propagation and all other optical elements in the beam path including a thin fused silica (FS) plano-convex lens (thickness ~ 2.2 mm) and an anti-reflection (AR) window (~ 2 mm). The laser pulses were focused via a fused-silica lens ($f$ = 75 cm) into the entrance of a 1.5 m long, neon-filled (13 bar) fused silica HCF with a bore diameter of 250 $\mu$m. The HCF was securely mounted on a V-groove aluminum holder to ensure mechanical and thermal stability during operation. Precise alignment was facilitated by a CCD-camera-based monitoring system at the capillary entrance. To ensure safe operation at neon gas pressure exceeding 20 bar, stainless steel ConFlat (CF) flanges were employed, sealed with a 2 mm AR coated entrance window and a 2 mm FS exit window placed at Brewster's angle (~ 55.4°). The total length of the host chamber was approximately 2.3 meters.

Using a beam profiler, we analyzed the near- and far-field beam profiles of the generated supercontinuum (SC) pulses right after collimation by a concave focusing mirror (silver coating, $f$ = 75 cm) placed downstream the HCF. The near-field measurements (Fig. 2(b)) were taken by recording the collimated beam, while the far-field data (Fig. 2(c)) were recorded using a concave mirror with a focal length of 60 cm. The low beam profile ellipticity of ~ 0.98 (near field) and ~ 1 (far field) indicate excellent beam quality.

The power of the pulses after beam collimation was 6 W, corresponding to an energy of ~ 600 $\mu$J per pulse, implying a module efficiency exceeding 65%. In next step, the pulses were directed into a compression set consisting of chirped mirrors (PC2151, UltraFast Innovations) optimized for the range from 500 nm to 1250 nm for ten reflections, along with a pair of FS wedges (2° 48'), allowing the fine-tuning of dispersion were used for the pulse compression. The overall transmission efficiency of the SC pulses through the dispersive mirror compressor was measured to be ~ 88%, resulting in an available pulse energy of approximately 530 $\mu$J for subsequent experiments.

For the accurate characterization of the spectral properties of our SC pulses at the exit of the HCF, a pre-calibrated dual-spectrometer module was employed. The module comprised an HR4000 (Ocean Optics) spectrometer, which covers the wavelength range from 250 nm to 950 nm, and an NIR Quest-512 (Ocean Optics) operating from 950 nm to 1400 nm. Spectral intensity calibration of the module was performed using a tungsten-deuterium light source (DH-3P-CAL, Ocean Optics), which provides a broad and stable reference spectrum. In this configuration, the SC spectrum was split and guided separately into the two branches of a bifurcated fiber to ensure a uniform intensity distribution before reaching the spectrometers. The spectrum was recorded at the exit of the

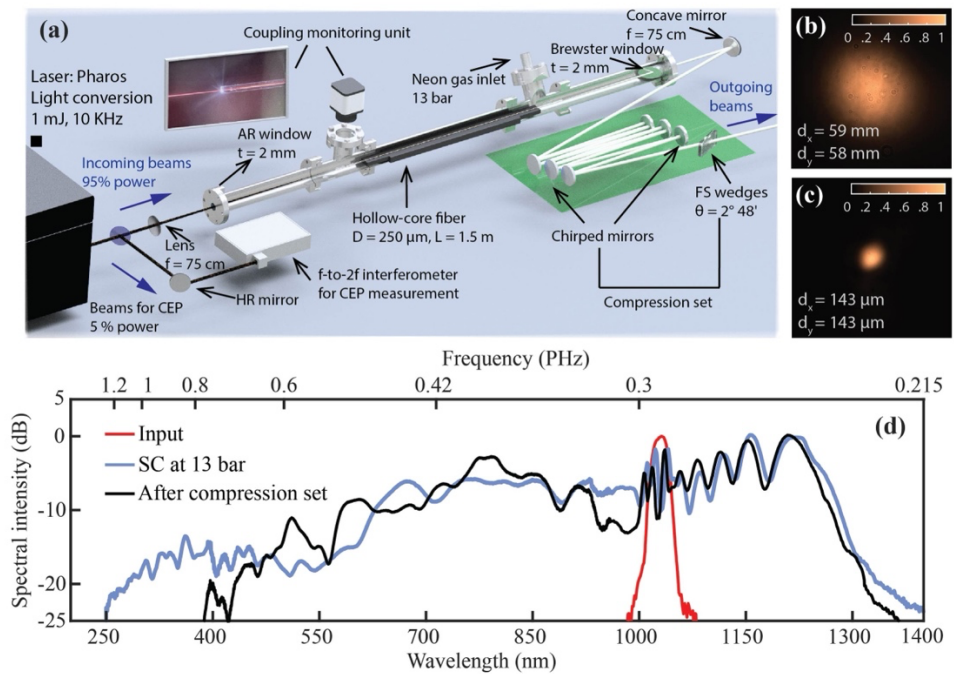

Fig. 2 (a) Experimental setup for spectral broadening and compression. (b)-(c) Beam profiles in the near-field and far-field, respectively. (d) Spectrum of the pulses at the exit of the fiber host chamber (blue line) and behind the chirped mirror compressor (black line) shown in (a). The red line denotes the spectrum of the pulse at the exit of the laser.

HCF host chamber via reflection off a specular reflectance standard (Ocean Optics) originally used for the intensity calibration of our dual-spectrometer module. Measurements are shown in Fig. 2(d). The SC (light blue line) is centered at ~ 1 $\mu$m and extends from 250 nm (1.2 Petahertz [PHz]) to 1400 nm (0.215 PHz) at the -25 dB intensity level, covering a bandwidth of ~1 PHz (2.4 octaves). This spectral bandwidth supports a FTL pulse duration of ~ 1.45 fs (~ 0.42 optical cycles). The spectrum after the chirped mirror compression module (black line in Fig. 2(d)) is limited by the reflectivity of the chirped mirrors and thus exhibits a narrower bandwidth, spanning from 400 nm to 1350 nm (~ 1.76 octaves at -25 dB), corresponding to a FTL pulse duration of ~ 2.85 fs.

To enable precise sampling of the electric field waveform of our pulses a Homochromatic Attosecond Streaking (HAS) setup was used [26, 27]. As illustrated in Fig. 3(a), the input optical pulse was split into a weak gate pulse (outer beam) and a high-intensity pump pulse (inner beam) using a dual-mirror assembly consisting of bare-aluminum-coated concave, concentric inner and outer mirrors ($f$ = 12.5 cm). An integrated imaging system based on a lens and a CCD camera ensured accurate spatial overlap of the pump and gate beams and their focusing onto the apex of an electrically grounded tungsten nanotip (apex radius ~ 35 nm). A piezo-based translation stage controlled the position of the inner mirror along the focusing direction providing attosecond precision in the timing between pump and gate pulses. Photoelectrons generated by the pump pulse from the tip apex were collected by a time-of-flight spectrometer whose entrance was positioned ~ 2 mm above the nanotip. Electron spectra were recorded as a function of pump-gate delay to compose HAS spectrograms.

A representative HAS spectrogram recorded using the compressed pulses is shown in Fig. 3(a). The white contour lines show the cut-off energy variation corresponding to the HAS vector potential $A_{HAS}$ of the gate pulse. The reconstructed electric field and instantaneous intensity

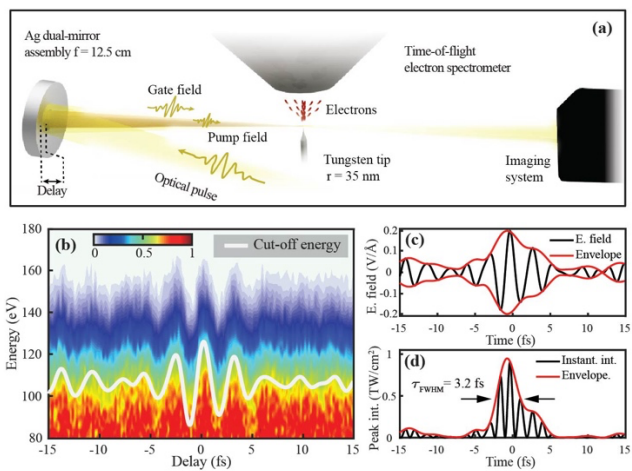

Fig. 3 (a) HAS setup for field sampling of the generated pulses. (b) HAS spectrogram of the pulses. The white curve marks the variation of the electron energy cut-off. (c) Electric field (black line) and its envelope (red line). (d) Instantaneous intensity profile (black line) and envelope (red line) of the pulse, as derived from the trace in (b).

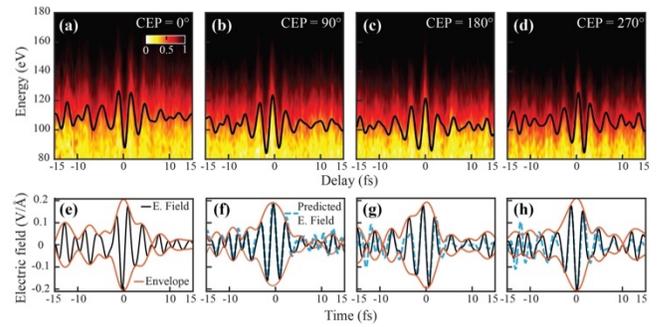

Fig. 4 HAS spectrogram of the pulse with CEP locked at: (a) 0°, (b) 90°, (c) 180°, and (d) 270° (false color). Electron energy cut-offs are marked by black lines. The corresponding electric fields are displayed in panels (e)–(h).

profiles are shown as black lines in Fig. 3(c) and 3(d), respectively. Fine-tuning of the wedge-pair allowed optimization of the pulse to a duration (FWHM) of 3.2 fs, with a central wavelength of 880 nm (~ 1.1 optical cycles). This result is nearing the FTL duration (FWHM) of 2.85 fs, as derived from the spectrum in Fig. 2(d) (black line), verifying the effectiveness of the compression strategy.

To demonstrate the capability of controlling the electric field waveforms of the pulses generated in our compression scheme, we systematically varied the CEP of our pulses and recorded HAS spectrograms. Figs. 4(a)–(d) display HAS spectrograms along with the evaluated HAS vector potentials (black lines) of the recorded pulses for CEP settings of 0°, 90°, 180°, and 270°, respectively. Figs. 4(e)–(h) present the reconstructed electric field waveforms, as retrieved from the data in Figs. 4(a)–(d). The dashed light-blue curves represent the theoretically predicted waveform shape, based on the field evaluated in Fig. 4(e) (marked as CEP = 0°), while solid black curves depict the measured electric field waveform at each CEP setting. The agreement between predicted and experimentally retrieved waveforms further validates the highly precise control of the generated waveforms.

In conclusion, we demonstrated single-stage super-octave (bandwidth > 1 PHz) broadening of mJ-level pulses generated from Yb:KGW amplifier and compression of a large portion of their spectrum into the single-cycle range (1.1 cycles). We achieved compression with a reasonable efficiency (> 65%) and exceptional spatial beam quality. The generated monocycles have a pulse duration (FWHM) of 3.2 fs, approaching the FTL value of 2.85 fs. Furthermore, the excellent agreement between the measured and theoretically predicted waveforms attests to the high stability and precision of the CEP control of our pulses. This cost-effective approach to tailored ultrafast waveform generation holds great promise for next generation ultrafast science studies and field-resolved metrology, as well as in attosecond science, strong-field physics, and high-resolution spectroscopy.

**Funding.** Funded by the Deutsche Forschungsgemeinschaft (DFG, German Research Foundation) - SFB 1477 "Light-Matter Interactions at Interfaces" (No. 441234705) and by the European Research Council (ERC) under the European Union's Horizon Europe research and innovation program (Advanced Grant Agreement No. 101098243).